%% file: sample-sigconf.tex
\documentclass[sigconf]{acmart}
\AtBeginDocument{%
  \providecommand\BibTeX{{%
    \normalfont B\kern-0.5em{\scshape i\kern-0.25em b}\kern-0.8em\TeX}}}

\copyrightyear{2023}
\acmYear{2023}
\setcopyright{acmlicensed}\acmConference[SIGIR '23]{Proceedings of the 46th International ACM SIGIR Conference on Research and Development in Information Retrieval}{July 23--27, 2023}{Taipei, Taiwan}
\acmBooktitle{Proceedings of the 46th International ACM SIGIR Conference on Research and Development in Information Retrieval (SIGIR '23), July 23--27, 2023, Taipei, Taiwan}
\acmPrice{15.00}
\acmDOI{10.1145/3539618.3591995}
\acmISBN{978-1-4503-9408-6/23/07}

\usepackage{multirow}
\usepackage{caption}
\usepackage{balance}
\usepackage{subcaption}
\usepackage[ruled,linesnumbered]{algorithm2e}

\begin{document}

\title{Neighborhood-based Hard Negative Mining for Sequential Recommendation}

\author{Lu Fan}
\email{cslfan@comp.polyu.edu.hk}
\affiliation{%
  \institution{The Hong Kong Polytechnic University}
  \city{Hong Kong S.A.R.}
  \country{China}
}

\author{Jiashu Pu}
\email{pujiashu@corp.netease.com}
\affiliation{%
  \institution{Fuxi AI Lab, NetEase Inc.}
  \city{Hang Zhou}
  \country{China}
}

\author{Rongsheng Zhang}
\email{zhangrongsheng@corp.netease.com}
\affiliation{%
  \institution{Fuxi AI Lab, NetEase Inc.}
  \city{Hang Zhou}
  \country{China}
}

\author{Xiao-Ming Wu}
\email{xiao-ming.wu@polyu.edu.hk}
\authornote{Corresponding author.}
\affiliation{%
  \institution{The Hong Kong Polytechnic University}
  \city{Hong Kong S.A.R.}
  \country{China}
}

\renewcommand{\shortauthors}{Lu Fan, Jiashu Pu, Rongsheng Zhang, \& Xiao-Ming Wu}

\begin{abstract}
    Negative sampling plays a crucial role in training successful sequential recommendation models. Instead of merely employing random negative sample selection, numerous strategies have been proposed to mine informative negative samples to enhance training and performance. However, few of these approaches utilize structural information. In this work, we observe that as training progresses, the distributions of node-pair similarities in different groups with varying degrees of neighborhood overlap change significantly, suggesting that item pairs in distinct groups may possess different negative relationships. Motivated by this observation, we propose a \emph{g}raph-based \emph{n}egative sampling approach based on \emph{n}eighborhood \emph{o}verlap (GNNO) to exploit structural information hidden in user behaviors for negative mining. GNNO first constructs a global weighted item transition graph using training sequences. Subsequently, it mines hard negative samples based on the degree of overlap with the target item on the graph. Furthermore, GNNO employs curriculum learning to control the hardness of negative samples, progressing from easy to difficult. Extensive experiments on three Amazon benchmarks demonstrate GNNO's effectiveness in consistently enhancing the performance of various state-of-the-art models and surpassing existing negative sampling strategies. The code will be released at \url{https://github.com/floatSDSDS/GNNO}.
\end{abstract}





\maketitle

\input{fig/BeautyHist}
\section{Introduction} \label{sec:intro}

Sequential recommendation (SR) is a task that predicts the next item that a user may be interested in given his or her past behavior sequence. Most SR models use Noise Contrastive Estimation (NCE), which brings positive items closer to the sequence and pushes negative ones away. Negative sampling strategies play a crucial role in the development of successful SR models. However, while the majority of SR models adopt uniform sampling strategies and focus on designing sequence encoders or incorporating additional training tasks, a few attempts have been made to focus on mining informative negative samples for SR. These attempts can be broadly categorized into two groups: (1) sampling-based methods~\cite{cho2014learning, gillick2019learning, wu2019scalable, karpukhin2020dense, xiong2020approximate} that aim to learn a proper distribution for negative mining, and (2) generation-based methods~\cite{wang2017irgan, yu2017seqgan, wang2018neural, huang2021mixgcf} that develop generative models that synthesize negative samples for model training.
Despite the various types of negative mining methods, to our knowledge, there is no effort that explicitly utilizes structural information concealed in user behavior sequences to mine hard negatives via graph analysis for SR. In this work, we develop a new negative sampling approach for SR by utilizing the structural information of behavior sequences and creating a negative sampler based on neighborhood analysis.

Our approach is motivated by a pilot experiment in Figure~\ref{fig:dist}, which visualizes the distributions of node-pair similarities in different groups at the different epochs during training an SR model. The nodes on a global weighted item transition graph (WITG) are divided into four groups w.r.t. the Jaccard similarity defined in Eq.~\ref{eq:jaccard} that measures the degree of neighborhood overlap between two nodes on the graph. Higher $J(i, j)$ indicates the neighborhoods of $i$ and $j$ have a larger overlap. 
Figure~\ref{fig:dist} shows that for each group, the distribution of embedding similarity shifts as training progresses. The distributions of \textit{group-zero} and \textit{group-low} exhibit similar patterns that they only change slightly. The two groups represent easy negatives, which, although less informative, are essential for training the SR model. 
Item pairs within \textit{group-medium} display some overlap in their respective neighborhoods. As training progresses, a significant shift in the distribution is observed. In this study, we treat item pairs in \textit{group-medium} as hard negative pairs, positing that they contribute more information during the training process.
Since item pairs in \textit{group-high} have strong connections and are likely to be false negatives, we propose to exclude them from the item pool for negative sampling.

Specifically, we propose a \emph{g}raph-based \emph{n}egative sampling approach based on \emph{n}eighborhood \emph{o}verlap (GNNO). GNNO first constructs a WITG as described in Sec~\ref{sec:witg}. Utilizing the built WITG, it selects negative samples for each target item  by considering the extent of the neighborhood overlap between the target item and any other item.
Additionally, GNNO employs curriculum learning (CL) to adjust the maximum hardness of negative samples, ranging from easy to hard.
Among existing graph-based negative sampling strategies, GNNO is most similar to RecNS~\cite{yang2022region} in that we also perform region division and propose sampling variations that account for unique characteristics in each region. The key differences between GNNO and RecNS are three-fold: 
(1) GNNO is designed for sequential recommendation and explicitly utilizes structural information in user behavior sequences;
(2) Rather than dividing the sampling region by $k$-hop distance, GNNO creates  a negative sampler that samples from a distribution based on the relative Jaccard index;
(3) GNNO suggests negative samples from distant regions are indispensable for training an SR model. In summary, the contributions of this paper include:
\begin{itemize}
\item To the best of our knowledge, this is the first work to investigate the structural properties of negative samples in SR. We observed that item pairs with varying levels of neighborhood overlap on WITG may exhibit distinct characteristics, signifying a valuable signal for hard negative discrimination in sequential recommendation.
\item We introduce GNNO, which adaptively samples negatives for each item based on the relative Jaccard similarity on WITG. 
We also employ CL to control the maximum hardness of negative samples during training. 
\item Comprehensive comparative experiments on three Amazon benchmarks demonstrate the effectiveness of our proposed method.
\end{itemize}

\input{fig/intuition}

\section{Related Works}

\paragraph{Negative Sampling Methods for Sequential Recommendation. }
As mentioned in Sec.~\ref{sec:intro}, attempts for informative negative sampling for SR are broadly categorized into two groups: sampling-based methods~\cite{cho2014learning, gillick2019learning, wu2019scalable, karpukhin2020dense, xiong2020approximate, wang2021cross} and generation-based methods~\cite{wang2017irgan, yu2017seqgan, wang2018neural, huang2021mixgcf}. ANCE~\cite{xiong2020approximate} proposes to sample hard negatives globally by using an asynchronously updated ANN index. CBNS~\cite{wang2021cross} proposes to sample negatives cross-batch other than simply in-batch sampling. SRNS~\cite{ding2020simplify} leverages the variance-based characteristics of false negatives and hard negatives to improve the sampling strategy.

\paragraph{Graph-based Negative Sampling Methods. }

Existing graph-based methods for negative sampling mostly concentrate on collaborative filtering (CF)-based recommendations~\cite{huang2021mixgcf, yang2022region}, graph contrastive learning~\cite{yang2020understanding, zhu2022structure, zhu2022structure, xia2022progcl}.
MixGCF~\cite{huang2021mixgcf} efficiently injects information of positive samples into negative samples via a mix-up mechanism. 
HORACE~\cite{zhu2022structure} and STENCIL~\cite{zhu2022structure} study heterogeneous graph contrastive learning and utilize global topology features including PageRank~\cite{langville2011google} and Laplacian vector. These features are usually static and would suffer from potential risks in static hard negative sampling. 
ProGCL~\cite{xia2022progcl} focuses on the problem of false negatives discrimination in graph contrastive learning. It proposes to solve the problem by fitting a two-component (true-false) beta mixture model to distinguish true negatives. 
RecNS~\cite{yang2022region} proposes the three-region principle to guide negative sampling for graph-based recommendation. 
It suggests to sample more negatives at an intermediate region while sampling less in adjacent and distant regions. 
In this work, we propose to construct a global weighted item transition graph to model sequential structural information for SR. To prevent the potential risks of static hard sampling, we propose to employ curriculum learning to adjust the maximum hardness of negative samples. 
Additionally, to address the false negative challenge~\cite{xia2022progcl} which is particularly severe in graph-based methods, we suggest to avoid sampling negatives from the ``adjacent region'' (w.r.t. the target item).

\section{Method}

    In this section, we start with the problem statement of sequential recommendation. Subsequently, we present our proposed approach. As illustrated in Figure~\ref{fig:method}, GNNO comprises three modules: (1) WITG construction; (2) overlapping-based negative sampling distribution generator; and (3) curriculum scheduler for negative sampling.

\subsection{Sequential Recommendation}
     
    Formally, let $\mathcal{U}$ and $\mathcal{I}$ denote the user and item sets from the interaction sequences, respectively. For each user $u\in \mathcal{U}$, we use a chronologically ordered list $\mathbf{s} = [i_1, i_2, ..., i_N]$ to denote his or her behavior sequence, where $i_t$ is the $t^{th}$ interaction item of $u$ and $N$ is the length of the sequence. The task of SR is to predict the subsequent user-item interactions with the given sequence $\mathbf{s}$.
     
    Training an SR model requires choosing an objective function $L$ and a negative sampling distribution $p_n$. 
    The commonly adopted Bayesian Personalized Ranking (BPR) loss~\cite{rendle2012bpr} w.r.t. all training sequences and time steps is defined as follows:
    \begin{equation}
        L_{BPR} = \sum_{(\textbf{s}_t, i_t)}{-\log \sigma(\hat{y}(\textbf{s}_t, i_t) - \hat{y}(\textbf{s}_t, i_t^-))}, \label{eq:bpr}
    \end{equation}
    where $\textbf{s}_t = [i_1, i_2, ..., i_{t-1}]$ is the historical sub-sequence of $\textbf{s}$ at time step $t$, $i_t$ is the target item, $\sigma$ denotes the sigmoid function, $\hat{y}(\mathbf{s}, i)$ is a trainable network that predicts a matching score between a sequence $\mathbf{s}$ and an item $i$. 
    $i_t^- \sim p_n$ is a negative item sampled from a distribution $p_n$. Optimizing Eq~\ref{eq:bpr} offers the model ability to score the target item $i_t$ higher than the negative item $i_t^-$ for a given $\mathbf{s}_t$. 
    
\subsection{Weighted Item Transition Graph (WITG)} \label{sec:witg}
    In contrast to the commonly used user-item graph, to make use of global information in behavior sequences, we propose to mine hard negatives on a WITG $\mathcal{G}(\mathcal{I}, \mathcal{E})$~\cite{zhang2022enhancing} where $\mathcal{E}$ is the edge set. 
    A WITG contains global item transition patterns extracted from all user behavior sequences in the training set $\mathcal{D}$. 
    $\mathcal{G}$ is constructed by traversing every sequence in $\mathcal{D}$. For a sequence $\mathbf{s} \in \mathcal{D}$, if there exists no edge between the items $i_m$ and $i_{(m+k)}$ in $\mathcal{G}$, we connect them and set the edge weight $w(i_m, i_{(m+k)})=1/k$, where $k$ represents the importance of a target item $i_m$ to its $k$-hop neighbor $i_{(m+k)}$ in $\mathbf{s}$. 
    Otherwise, if there is already an edge between them, we update the edge weight as $w(i_m, i_{(m+k)}) \leftarrow  w(i_m, i_{(m+k)}) + 1/k$. 

\subsection{Neighborhood-based Negative Sampler}

    Given a target item, we can divide the other items into groups w.r.t. their neighborhood overlap with the target item. As analyzed in Sec.~\ref{sec:intro}, these groups demonstrate distinct characteristics during the training of a SR model, thus it can be advantageous to differentiate and treat them differently.   
    Here, we propose to use the Jaccard similarity to measure the degree of neighborhood overlap between the target item and the others. 
    Let $\mathcal{N}(i)$ denote the neighbor set of an item $i$ on $\mathcal{G}$. The Jaccard similarity between any two items $i$ and $j$ can be defined as:
    \begin{equation}
        J(i, j) = \frac{\mathcal{N}(i) \cap \mathcal{N}(j)}{\mathcal{N}(i) \cup \mathcal{N}(j)}. \label{eq:jaccard}
    \end{equation}  
    If $i$ and $j$ have many common neighbors, they will have a large similarity. Meanwhile, the denominator acts as a normalizer, and the similarity will be small if the degree of either $i$ or $j$ is large.

    As explained in Figure~\ref{fig:dist}, \textit{group-medium} is likely to be quality hard negatives, hence we want to give them higher sampling weight. Meanwhile, we want to keep \textit{group-zero} and \textit{group-medium} for diversity, but we want to give them low sampling probability. Finally, we want to exclude \textit{group-high} as they are likely to be false negatives.
    Therefore, we propose to sample the negatives for a target item $i$ from the following distribution:
    \begin{equation}
        p_n(i^- | i) = \frac{e^{J(i, i^-)}}{\sum_{j \in \mathcal{N}'(i)} e^{J(i, j)}}, 
        \label{eq: pn}
    \end{equation}
    where $\mathcal{N}'(i)=\mathcal{I} \backslash \{j|J(i,j) > \lambda\}$ is the item set without \textit{group-high}, and $\lambda$ is 
    the threshold for \textit{group-high} at the current training step.

\subsection{Curriculum Scheduler for Negative Sampling}
    Concentrating too much on hard negatives in the early training stage may bring negative effects to the model~\cite{castells2020superloss}. 
    We therefore employ curriculum learning (CL) techniques to schedule the hardness of the negatives. The main idea of CL is to order negative samples during training based on their difficulty~\cite{wang2021survey}. 
    Let $Q$ be the maximum training step, then for each training step $q\in [1, ..., Q]$, we update $\lambda$ to form $\mathcal{N}'(i)$ in Eq.~\ref{eq: pn}.
    The maximum Jaccard score is updated from an initial hardness $b$ with a linear pacing function $f(q)$:
    \begin{equation}\label{eq:CL}
        f(q) = c * q + b,
    \end{equation}
    where $q$ is the current time step and $c$ is the pace coefficient. For simplicity, we set $b=0$ in our experiments. 
    Moreover, we set a maximum hardness $\lambda_{max}$ to clip the growth of $\lambda$. Finally, $\lambda$ is updated as:
    \begin{equation}\label{eq:CL2}
        \lambda(q) = min(f(q), \lambda_{max}).
    \end{equation}

\section{Experiments}
\subsection{Experimental Setup}
\subsubsection{Datasets}

    Our experiments are conducted on three subsets of the well-known Amazon dataset\footnote{http://jmcauley.ucsd.edu/data/amazon}~\cite{amazon/ni2019justifying}, 
    which includes rich user-item review interactions. Specifically, we choose the sub datasets \textit{Beauty, Phone, and Toy} for our empirical study. In our experiments, we use the 5-core data. Items and users with no positive records are filtered out for each subset. The statistics of the filtered datasets are shown in Table~\ref{tb: dataset}.
    
\subsubsection{Baselines}

\input{table/dataset.tex}
\input{table/parameters.tex}

    We compare our method with both state-of-the-art sequential recommendation methods and negative sampling strategies. As shown in Table~\ref{tb:rst_rec}, we compare our method with the following baselines for sequential recommendation: \textbf{BPRMF}~\cite{rendle2012bpr}, \textbf{Caser}~\cite{tang2018personalized}, \textbf{GRU4Rec}~\cite{hidasi2015session}, \textbf{BERT4Rec}~\cite{sun2019bert4rec}, \textbf{SASRec}~\cite{kang2018self}, 
    \textbf{TimiRec} \cite{wang2022target},
    and \textbf{ContraRec}~\cite{wang2023sequential}.
    Meanwhile, as shown in Table~\ref{tb:rst_hn}, we also compare our method with the following negative sampling approaches: 
    \textbf{DNS}~\cite{zhang2013optimizing} that adaptively samples the negative item scored highest by the recommender, 
    \textbf{MCNS}~\cite{yang2020understanding}that samples negatives with the distribution sub-linearly related to the positive distribution and accelerates the sampling process by Metropolis-Hastings, and \textbf{MixGCF}~\cite{huang2021mixgcf} that injects positive samples into negatives via hop mixing to synthesize hard negatives. Note that MCNS and MixGCF are graph-based but DNS is not.
    
\subsubsection{Implementation details}
    All recommendation baselines are implemented with ReChorus\footnote{https://github.com/THUwangcy/ReChorus}, a framework for top-$k$ recommendation. GNNO and other negative sampling strategies are implemented on the state-of-the-art SR method ContraRec. We use the default settings of ContraRec.
    Specifically, we use Adam as the optimizer and BERT4Rec as the sequence encoder. 
    The batch size is set to $4096$, and the dimension of hidden units is set to $64$. 
    The hyper-parameters of GNNO, as shown in Table~\ref{tb:param}, are selected based on the performances on the validation set using grid search.

\subsubsection{Evalutaion Protocals} 

    Following \cite{wang2023sequential}, we adopt the commonly used leave-one-out strategy to evaluate model performance. 
    We employ two evaluation metrics: \textbf{hit rate (HR@K)} and \textbf{normalized discounted cumulative gain (NDCG@K)}. 
    For each sequence, the target products are mixed up with candidates randomly sampled from the entire product set, forming a candidate set of size $1000$.

\subsection{Results and Analysis}

\subsubsection{Comparisons with Negative Sampling Baselines}
\input{table/results_hn}

Table~\ref{tb:rst_hn} summarizes the recommendation performance of GNNO in comparison to existing negative sampling approaches on three Amazon benchmarks. GNNO outperforms the baselines in almost all cases, demonstrating its efficacy. It can be seen that all graph-based methods perform better than DNS, which highlights the advantages of incorporating structural information for negative sampling.

\subsubsection{Comparisons with Baselines for Sequential Recommendation}

\input{table/results}
Table~\ref{tb:rst_rec} shows the recommendation performance of GNNO against state-of-the-art methods for SR on three Amazon benchmarks. 
GNNO improves over ContraRec on every dataset. It demonstrates the effectiveness of the proposed negative sampling strategy, showing that pushing away negatives with some neighborhood overlap is beneficial for SR.
In addition, both ContraRec and GNNO outperform other baselines consistently, where the performance gain is probably brought by the context-context contrastive learning task.

\section{Conclusion}

    In this work, we observed that as training progresses, the embedding similarity between item pairs in different groups with varying degrees of neighborhood overlap on a weighted item transition graph (WITG) changes significantly. Based on this observation, we propose GNNO, which samples negatives with respect to the Jaccard index on a global WITG. Additionally, GNNO employs curriculum learning to manage the hardness of negative samples at each training step. Extensive experiments on three Amazon benchmarks demonstrate the effectiveness of our proposed method. In future work, we plan to study hard negative mining over dynamic graphs for sequential recommendation.

\section{Acknowledgement}

We would like to thank the anonymous reviewers for their helpful
comments. This research was partially supported by the General Research Fund No.15222220 funded by the UGC of Hong Kong.

\bibliographystyle{ACM-Reference-Format}
\balance
\bibliography{sample-base}

\end{document}

%% file: fig/BeautyHist.tex
\begin{figure}
\centering

\scalebox{1}{
\begin{subfigure}{.23\textwidth}
  \centering
  \includegraphics[width=\linewidth]{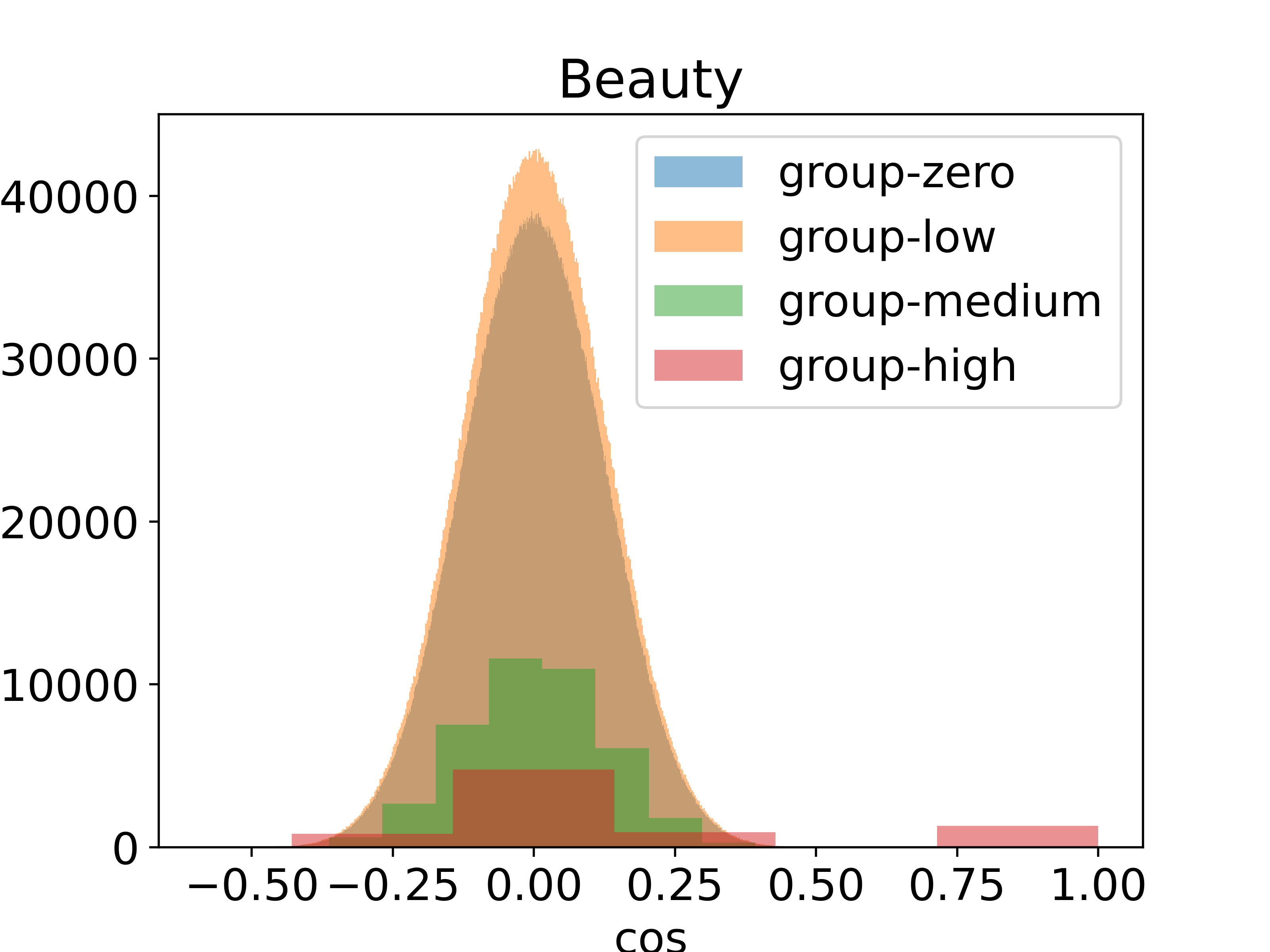}
\end{subfigure}%
\begin{subfigure}{.23\textwidth}
  \centering
  \includegraphics[width=\linewidth]{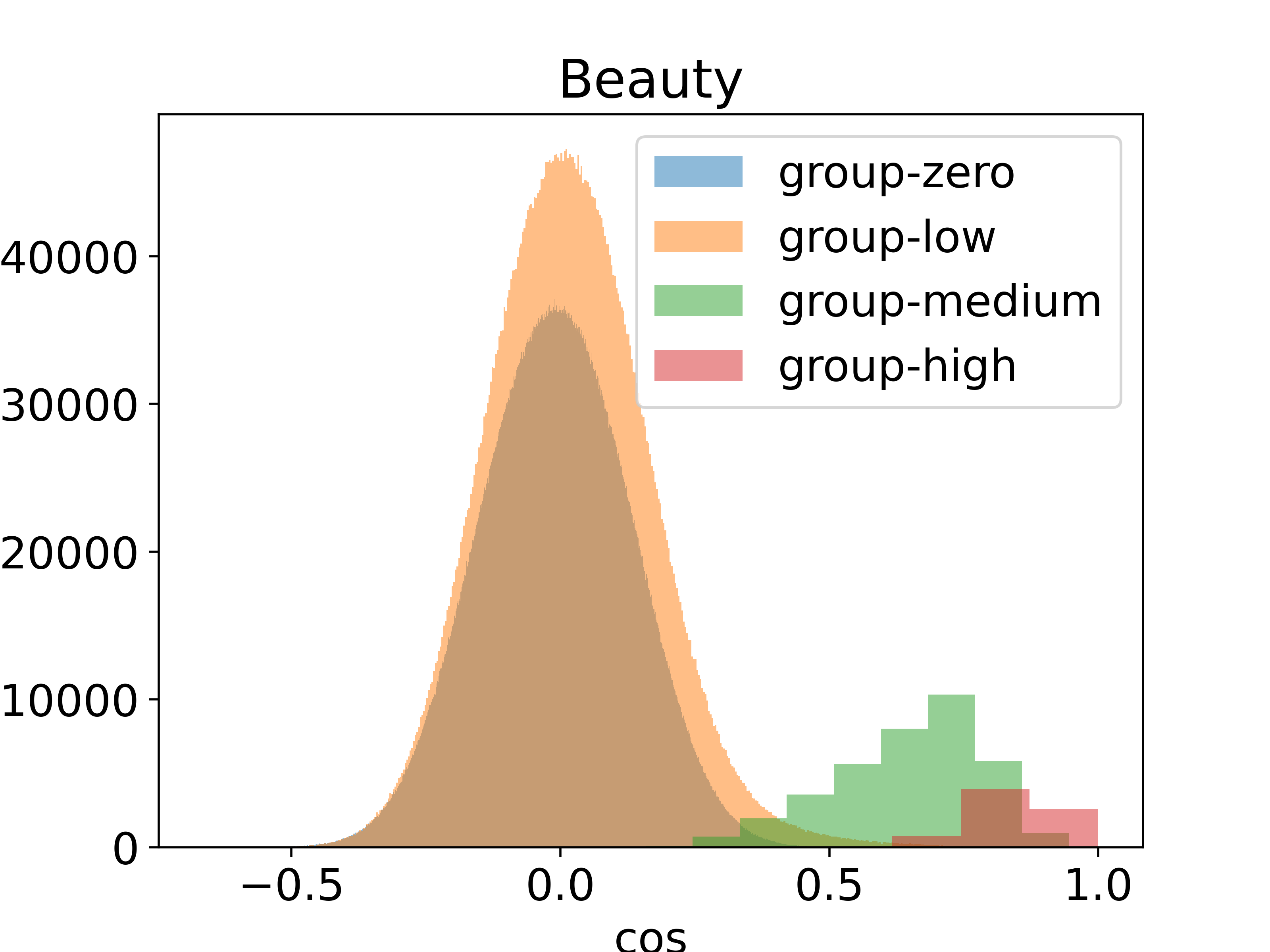}
\end{subfigure}
}
\caption{Visualization of the distributions of node-pair similarities in different groups on the \textit{Beauty} dataset at the training epoch 0 and 20, respectively. 
Nodes are divided into four groups w.r.t. their Jaccard similarity $J(i, j)$ on a global weighted item transition graph. 
Node/item pairs in \textit{group-zero} have no common neighbors, those in \textit{group-low} have $J(i, j)\in(0, 0.15]$, those in \textit{group-medium} have $J(i,j)\in(0.15, 0.3]$, and those in \textit{group-high} have $J(i,j)\in (0.3, 1.0]$. 
Note that due to the considerable differences in group size, we use different bin values for different groups.
} 
\label{fig:dist}
\end{figure}

%% file: fig/intuition.tex
\begin{figure}
    \centering
    \includegraphics[width=\linewidth]{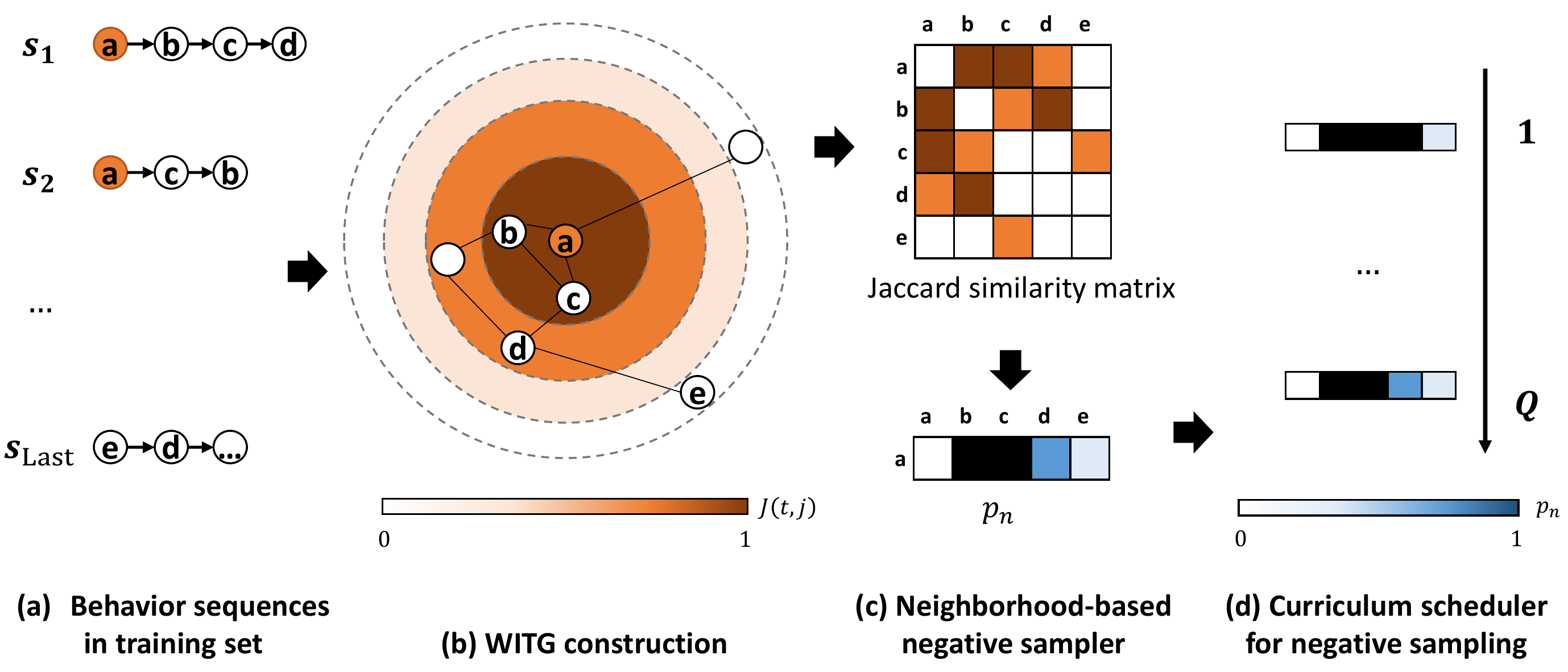}
    \caption{Illustration of our proposed GNNO framework.
    The false negatives for a given target node are annotated as black blocks. For example, for the target node $a$, nodes $b$ and $c$ are false negatives and hence excluded from the sampling distribution for $a$.
    }
    \label{fig:method}
\end{figure}

%% file: table/dataset.tex
\begin{table}[]
\small
\caption{Dataset statistics.}
\scalebox{0.8}{
\begin{tabular}{lccccc}
\toprule
\textbf{} & \#user  & \#item  & \#entry  & \#edge   & \multicolumn{1}{c}{sparseness} \\ \hline
Beauty    & 22,363 & 12,101 & 198,502 & 530,266 & 20.01\%                             \\
Toys      & 19,413 & 11,925 & 148,455 & 486,740 & 16.32\%                             \\
Phone     & 22,364 & 12,102 & 176,139 & 530,266 & 24.55\%                             \\ 
\bottomrule
\end{tabular}
}
\label{tb: dataset}
\end{table}

%% file: table/parameters.tex
\begin{table}[]
\small
\caption{Hyper-parameter settings. \#$neg_{hard}$ and \#$neg_{rand}$ indicate the number of hard negatives and random negatives, respectively. }
\scalebox{0.8}{
\begin{tabular}{lccccc}
\toprule
& \#$neg_{hard}$ & \#$neg_{rand}$ & $c$ (Eq.\ref{eq:CL})    & $\lambda_{max}$ (Eq.\ref{eq:CL2}) \\ \hline
Beauty                         & 9    & 16   & 0.04 & 0.5\\
Toys               & 2    & 10   & 0.05 & 0.2\\
Phones & 4    & 10   & 0.01 & 0.9\\ 
\bottomrule
\end{tabular}
}
\label{tb:param}
\end{table}

%% file: table/results_hn.tex
\begin{table}[]
\small
\caption{Comparison of our proposed method with hard negative mining baselines on three Amazon review sub datasets. The best results are highlighted in bold.}
\scalebox{0.8}{
\begin{tabular}{clllll}
\toprule
\hline
\multicolumn{1}{l}{\textbf{Dataset}} & \textbf{Method}  & \textbf{HR@5}   & \textbf{NDCG@5} & \textbf{HR@20}  & \textbf{NDCG@20} \\ \hline
\multirow{4}{*}{\textbf{Beauty}}     & \textbf{DNS}     & 0.3858          & 0.2984          & 0.5890          & 0.3561           \\
                                     & \textbf{MCNS}    & 0.4101          & 0.3141          & 0.6135 & 0.3721           \\
                                     & \textbf{MixGCF}  & 0.4123          & 0.3168          & 0.6140          & 0.3743           \\ \cline{2-6} 
                                     & \textbf{GNNO}   & \textbf{0.4149} & \textbf{0.3215} & \textbf{0.6174} & \textbf{0.3792}  \\ \hline

\multirow{4}{*}{\textbf{Phone}}      & \textbf{DNS}     & 0.4307          & 0.3300          & 0.6563          & 0.3943           \\
                                     & \textbf{MCNS}    & 0.4794          & 0.3646          & 0.7179          & 0.4334           \\
                                     & \textbf{MixGCF}  & 0.4810          & 0.3655          & 0.7197          & 0.4342           \\ \cline{2-6} 
                                      & \textbf{GNNO}   & \textbf{0.4897} & \textbf{0.3754} & \textbf{0.7249} & \textbf{0.4430}  \\ \hline
\multirow{4}{*}{\textbf{Toy}}        & \textbf{DNS}     & 0.3470          & 0.2666          & 0.5598          & 0.3266           \\
                                     & \textbf{MCNS}    & 0.4035          & 0.3084          & 0.6238          & 0.3711           \\
                                     & \textbf{MixGCF}  & 0.4049          & 0.3097          & \textbf{0.6252}          & 0.3723           \\ \cline{2-6} 
                                     & \textbf{GNNO} & \textbf{0.4074} & \textbf{0.3148} & 0.6245 & \textbf{0.3762}  \\ \hline
 \bottomrule
\end{tabular}
}
\label{tb:rst_hn}
\end{table}

%% file: table/results.tex
\begin{table}[]
\small
\caption{Comparison of our proposed method with baselines for sequential recommendation on three Amazon review sub-datasets. The best results are highlighted in bold.}
\scalebox{0.8}{
\begin{tabular}{llllll}
\toprule
\hline
\textbf{Dataset}                 & \textbf{Method}    & \textbf{HR@5}   & \textbf{NDCG@5} & \textbf{HR@20}  & \textbf{NDCG@20} \\ \hline
\multirow{8}{*}{\textbf{Beauty}} & \textbf{BPRMF}     & 0.3588          & 0.2593          & 0.5716          & 0.3202           \\
                                 & \textbf{Caser}     & 0.3198          & 0.2238          & 0.5772          & 0.2970           \\
                                 & \textbf{GRU4Rec}   & 0.3254          & 0.2318          & 0.5762          & 0.3030           \\
                                 & \textbf{BERT4Rec}  & 0.3590          & 0.2658          & 0.5734          & 0.3266           \\
                                 & \textbf{SASRec}    & 0.3653          & 0.2780          & 0.5744          & 0.3372           \\
                                 & \textbf{TimRec}    & 0.3781          & 0.2812          & 0.5958          & 0.3433           \\
                                 & \textbf{ContraRec} & 0.4112          & 0.3158          & 0.6111          & 0.3727           \\ \cline{2-6} 
                                 & \textbf{GNNO}   & \textbf{0.4149} & \textbf{0.3215} & \textbf{0.6174} & \textbf{0.3792}  \\ \hline
\multirow{8}{*}{\textbf{Phone}}  & \textbf{BPRMF}     & 0.3690          & 0.2709          & 0.5945          & 0.3352           \\
                                 & \textbf{Caser}     & 0.3873          & 0.2745          & 0.6727          & 0.3560           \\
                                 & \textbf{GRU4Rec}   & 0.4122          & 0.2973          & 0.6935          & 0.3777           \\
                                 & \textbf{BERT4Rec}  & 0.4209          & 0.3145          & 0.6664          & 0.3849           \\
                                 & \textbf{SASRec}    & 0.4432          & 0.3349          & 0.6809          & 0.4032           \\
                                 & \textbf{TimRec}    & 0.4338          & 0.3241          & 0.6712          & 0.3925           \\
                                 & \textbf{ContraRec} & 0.4831          & 0.3673          & 0.7210          & 0.4358           \\ \cline{2-6} 
                                 & \textbf{GNNO}   & \textbf{0.4897} & \textbf{0.3754} & \textbf{0.7249} & \textbf{0.4430}  \\ \hline
\multirow{8}{*}{\textbf{Toys}}   & \textbf{BPRMF}     & 0.3107          & 0.2255          & 0.5255          & 0.2864           \\
                                 & \textbf{Caser}     & 0.2921          & 0.2000          & 0.5584          & 0.2757           \\
                                 & \textbf{GRU4Rec}   & 0.3113          & 0.2168          & 0.5802          & 0.2934           \\
                                 & \textbf{BERT4Rec}  & 0.3457          & 0.2561          & 0.5582          & 0.3164           \\
                                 & \textbf{SASRec}    & 0.3594          & 0.2731          & 0.5694          & 0.3328           \\
                                 & \textbf{TimRec}    & 0.3535          & 0.2628          & 0.5849          & 0.3287           \\
                                 & \textbf{ContraRec} & 0.4013          & 0.3065          & 0.6177          & 0.3676           \\ \cline{2-6} 
                                 & \textbf{GNNO} & \textbf{0.4074} & \textbf{0.3148} & \textbf{0.6245} & \textbf{0.3762}  \\ \hline
\end{tabular}
}

\label{tb:rst_rec}
\end{table}